\documentclass[twocolumn,prl,showpacs]{revtex4}

\usepackage{graphicx}
\usepackage{dcolumn}
\usepackage{bm}
\usepackage{amssymb}
\usepackage{amsmath}

\begin{document}
\title{
Temporal and dimensional effects in evolutionary graph theory
}

\author{C. J. Paley$^\dagger$, S. N. Taraskin$^*,\dagger$, 
 and S.R. Elliott$^\dagger$ } 
 
\affiliation{$^\dagger$Department of Chemistry, 
University of Cambridge, 
Lensfield Road, Cambridge CB2 1EW, UK}

\affiliation{$^*$St. Catharine's College, 
 Cambridge CB2 1RL, UK}

  


\date{\today}

\begin{abstract}
The spread in time of a mutation through a population
 is studied analytically and computationally in fully-connected networks and on spatial lattices.
The time, $t_*$, for a favourable mutation to dominate scales with population size $N$ as $N^{(D+1)/D}$ in 
 $D$-dimensional hypercubic lattices and as $N\ln N$ in fully-connected graphs. 
It is shown that the
 surface of 
the interface between mutants and non-mutants is crucial in predicting the dynamics of the system.
Network topology has a significant effect on the equilibrium
 fitness of 
a simple population model incorporating multiple mutations and sexual
 reproduction. 
Includes supplementary information.
\end{abstract}


\maketitle
Population models have been used to study various problems 
in evolutionary biology \cite{Baake_00, Hartl:book, Ewens:book}.
Generally, these models assume a non-spatial population 
\cite{Kimura_64, Kimura_66, Kondrashov_88, Cohen_05}.
As competition in the wild is generally local and spatial \cite{Tilman:book}, 
the impact of this assumption needs to be examined carefully.
There is evidence that spatially subdividing populations can change 
the dynamics of alleles spreading through a population 
\cite{Whitlock_03, Maruyama_70, Slatkin_81}.
Spatial structure has also been shown to change the dynamics of  
co-operative games\cite{Nowak_93, Hauert_04}. 

Recently, Lieberman et al.\cite{Lieberman_05} introduced 
'evolutionary graph theory' to allow the effects of spatial structure 
on evolutionary dynamics to be studied directly, 
placing Moran's process\cite{Moran_58} 
on a network.
On a $2D$-network, this model is equivalent 
to the Williams-Bjerknes model \cite{Williams_72}.
Lieberman et al. showed that, in networks characterized by a 
 symmetric stochastic matrix, 
mutants have a fixation probability that depends on the mutant
fitness, 
$r$ (defined relative to a background wild-type fitness of unity),  
and on the number of nodes in the system, $N$, but which is the same for 
all members of that class of networks.  
Other networks, such as scale-free networks, 
were found to lead to different fixation probabilities \cite{Lieberman_05}.

The main aim of this paper is to investigate analytically and numerically 
the \textit{dynamics} of evolution in 
networks with symmetric stochastic matrices. 
We demonstrate that 
the mutant fixation time significantly depends on the network topology and 
dimensionality 
(as well as $N$ and $r$),
despite the fact that the fixation probability in all these 
networks is the same in the limit of infinite time.  
In a related model, allowing for simultaneous multiple mutations,
 the equilibrium fitness is defined by a balance equation with 
a typical time much less than the fixation time and hence 
the equilibrium fitness of a population model  
depends on the Euclidean dimensionality of the network. 
Thus,  
approximating realistic spatial networks  by fully-connected graphs 
results in an appreciable overestimate 
of the equilibrium fitness.  
We also demonstrate that the topology of the network affects differently 
the equilibrium fitness of sexual and asexual populations.    

In 'evolutionary graph theory' \cite{Lieberman_05}, each vertex of a graph 
represents an individual with fitness $r$ for a mutant and $1$ for a wild-type organism.
On each turn (time step), an individual on a node $i$ is selected for reproduction, 
with a probability linearly proportional to its fitness.
A clone of the selected individual is placed onto one of the nodes, $j$,  
connected to it by the network which 
is chosen with a probability, $w_{ij}$,  given by the stochastic matrix. 
%
The individual previously at $j$ is replaced, 
conserving the population size.
Fixation of the mutant gene occurs when the number of mutants, $m$, 
reaches $N$, and extinction occurs when $m=0$.

Here, we consider the special class of graphs with 
fixed node connectivity, $Z$, and
$w_{ij}=w_{ji}=1/Z$ for connected nodes and 
$w_{ij}=0$ otherwise,
 and analyse the spread of 
an advantageous mutation
($r>1$).

The number of mutants, $m(t)$, 
increases, decreases or remains unchanged in each time step with
probabilities,  $p_{\uparrow}$,  $p_{\downarrow}$, and  
$1- p_{\uparrow}-p_{\downarrow}$, respectively. 
Changes occur at the interface consisting of $A$ links between mutants 
and wild-types. 
The value of $p_{\uparrow}$ is given by the probability that a given
mutant reproduces in the turn, $p_{m,\text{repr}}=r/[mr+(N-m)\times 1]$,
multiplied by the number of mutants and 
$p_{m,\text{inter}}=A/(mZ)$, which is the probability that the mutant passes
its offspring across the interface, 
i.e. $p_{\uparrow}=(A/Z)r/[mr+(N-m)]$. 
Similarly,
$p_{\downarrow}=(A/Z)\times 1/(mr+(N-m))$. 
Taking the deterministic (fluctuations are ignored), continuous-time 
and continuous-mutant population
number  approximation ($1\ll m \ll N$), we obtain the following 
 equation \cite{Paley:supp} 
\begin{equation}
 \frac{\text{d}m}{\text{d}t}\simeq R_{\uparrow}-R_{\downarrow} = \frac{A}{Z}\frac{r-1}{mr+(N-m)}~,
\label{eq:master}
\end{equation}
where $R_{\uparrow(\downarrow)}= p_{\uparrow(\downarrow)}/\Delta t$ 
(with $\Delta t=1$ being the simulation time step length),
i.e. the change in the mutant population per time step is, on average, equal
to the 
probability that the population grows minus the probability that it 
decreases.

The number of links between mutants and non-mutants may vary with $m(t)$, 
and 
for some networks, the functional form of $A(m)$ is known exactly. 
For example, in the linear chain ($Z=2$, periodic boundaries), 
$A=2$ for $m < N$ and $A=0$ for $m=N$, and hence the 
solution of Eq.~(\ref{eq:master}), with $m(0)=1$, is straightforward, 
\begin{equation}
m(t)=-\frac{N}{r-1}+\sqrt{\left(\frac{N}{r-1}+1\right)^2+2t}~,
\label{e2}
\end{equation}

for $t\leq t_{\text{f}}=N^2(r+1)/(r-1)$, 
where the upper limit $t_{\text{f}}$ is the fixation time which scales
quadratically with $N$. 

\begin{figure}[ht]
\includegraphics[angle=270, width=0.49\textwidth]{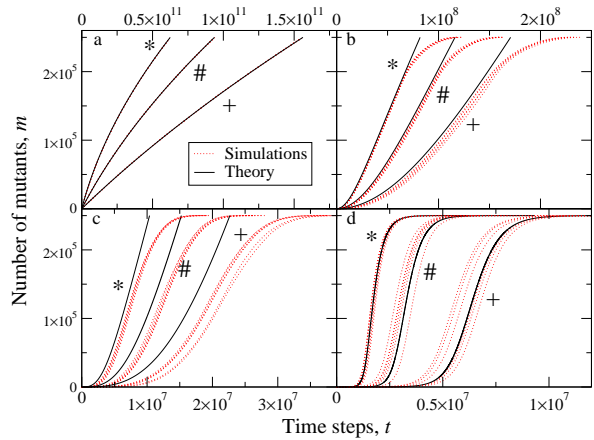}
\caption{
\label{f1}(Color online) 
Number of mutants, $m$, vs. time, $t$, in the models defined on networks 
of different topologies: (a) linear chain, (b) square lattice, (c) cubic lattice and 
(d) fully-connected graph. 
Predictions (solid black line) are compared to data from six simulation runs (dashed red line).
The plots marked '$*$' are for mutants of $r=3$, '$\#$' corresponds to $r=2$, and '$+$' to $r=1.5$. 
$N=250,000$ has been used in simulations 
for all networks except (c) with $N=250,047$. 
The theoretical plot in (a) overlays the simulation data.
}
\end{figure}

In the case of a fully-connected graph ($Z=N-1$), $A=m(N-m)$, and the rate 
equation can be solved implicitly, resulting in (for $m(0)=1$):
\begin{equation}
t=\frac{(N-1)}{(r-1)}
\ln\left[\frac{(N-1)^{r}m}{(N-m)^{r}}\right]~.
\label{e3}
\end{equation}
Simulations support these analytical expressions 
(see Fig.~\ref{f1}(a) and (d)). 
The continuum approximation used in Eq.~(\ref{eq:master}) breaks down for 
$m\to 0$ or $m \to N$. 
This causes large fluctuations at the initial stage of invasion, which lead to 
displacements of $m(t)$ along the time axis 
(see Fig.~\ref{f1}), 
being equivalent to variations
in the initial condition. 
The shape of the curve does not change and agrees well 
with the analytical predictions. 
It might be biologically important that the size of the initial fluctuations 
 increases with decreasing mutant fitness \cite{Rozen_02}, as can be clearly 
seen from Fig.~\ref{f1}(d). 
The fluctuations, which separate the curves of each simulation run are of greater consequence in the high-dimension systems where the overall time-scales are shorter.

For square ($Z=4$) and cubic ($Z=6$) lattices, the relationship between 
$A$ and $m$ is multi-valued and unknown. 
The current belief is that the mutant/non-mutant interface is a self-affine 
surface with 
\begin{equation}
A = \beta m^{\gamma}~, 
\label{e4}
\end{equation}
and $\gamma = (D-1)/D$ \cite{Meakin_93}, which is consistent with our numerical
data  (see Table~\ref{table1}).     
%
%
\begin{table}
\caption{\label{table1}
Values for $\gamma$ found from fits to Eq.~(\ref{e4}) with
16 realisations used for each fit. 
In the 2D case, fits were performed with between 60,000 and 150,000 mutants; 
in the 3D case, between 45,000 and 75,000 mutants were used. 
}
\begin{ruledtabular}
\begin{tabular}{lcr}
Mutant fitness, $r$&Square lattice, $\gamma$&Cubic lattice, $\gamma$\\
\hline
1.5 & $0.513 \pm 0.014$ & $0.684 \pm 0.012$ \\
2 & $0.504 \pm 0.008$ & $0.700 \pm 0.012$ \\
3 & $0.512 \pm 0.008$ & $0.692 \pm 0.012$ \\
100 & $0.509 \pm 0.006$ & $0.698 \pm 0.007$ \\
\end{tabular}
\end{ruledtabular}
\end{table}

Using the numerical estimates for $\beta$ and 
$\gamma= (D-1)/D$,
the rate Eq.~(\ref{eq:master}) can be solved implicitly for 
$2D$- and $3D$-lattices ($m(0)=1$),  
\begin{equation}
t=\frac{Z}{\beta}\left(\frac{1}{2-\gamma}\left(m^{2-\gamma}-1\right)
+\frac{N}{(1-\gamma)(r-1)}\left(m^{1-\gamma}-1\right)\right)~. 
\label{e5}
\end{equation}
Numerical simulations are in better agreement with analytical predictions 
in the case of $2D$- than $3D$-lattices. 
Both predictions fail when edge effects (above the inflexion points in Figs.~\ref{f1} 
(b, c) for simulations) become important, i.e. at sufficiently
large times. 
The systematic deviation, overestimating the mutant spread speed, of the simulated data from analytical predictions 
is due to the violation of Eq.~(\ref{e4}) (see Table~\ref{table1}) for small times, when the number of mutants is small and one is far from the asymptotic regime. 
This effect is more significant in the cubic lattice, which 
tallies with work on Eden growth exhibiting more pronounced crossover effects 
 in higher dimensions \cite{Devillard_89}. 
These deviations are also more substantial when mutants have a smaller
comparative fitness (cf. curves for different values of $r$ in
Fig.~\ref{f1} (b, c)). 
Real favourable mutations generally confer an advantage far smaller than
50\% (corresponding to $r=1.5$ in Fig.~\ref{f1}) \cite{Rozen_02}, and 
thus this discrepancy, and early fluctuations, 
are potentially of importance in
biological systems.
The less-biased errors from the approximations in Eq.~(\ref{eq:master}) are
present in all the simulations, 
and, for a given $r$, are of similar magnitude in time - the spread of data in
time is of 
a similar magnitude in the body of the plot for all the system types, 
but shows up most in fully-connected systems which exhibit the shortest time to fixation.


\begin{figure}[ht]
\includegraphics[angle=270, width=0.49\textwidth]{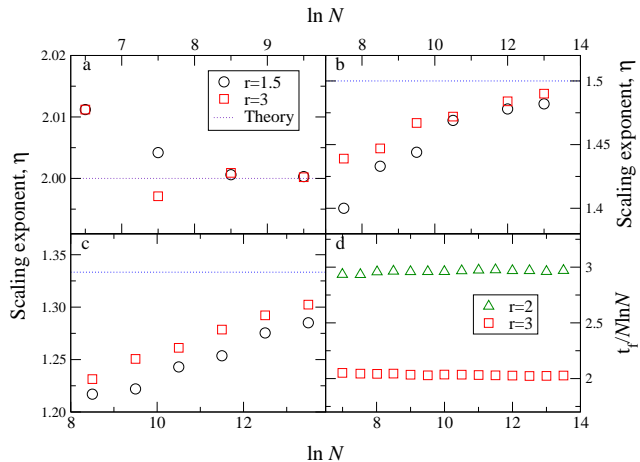}
\caption{
\label{f2}(Color online) 
Relationship between scaling exponents for fixation time and population size:
(a) Linear chain;
(b) Square lattice;
(c) Cubic lattice;
(d) Fully-connected graph.
For (a)-(c), theory predicts $t_f \propto N^{\eta}$.  
Using simulations for system sizes spaced at $\ln N=0.5$ intervals, least-square fits were performed over three adjacent times to obtain $\eta$.
Each point is therefore an approximation to the gradient of a $\ln t_f$ \textit{vs}  $\ln N$ plot at the point.
The predicted scaling ($\eta = D/(D+1)$) is shown with dashed lines.
For (d), the theory predicts that the time to fixation on a fully-connected graph divided by $N \ln N$ will be constant.
Between 100 and 3300 data points were averaged for each value of $t_f$.
}
\end{figure}

It follows from Eqs.(2)-(5)  that the typical time, $t_*$, taken for a mutant to dominate the graph 
(i.e. to occupy a large fraction of sites, $m(t_*) \sim N$) 
depends strongly on the network topology.
In a fully-connected graph, $t_*$ scales as $N\ln N$, and, 
in a $D$-dimensional Euclidian network, as $N^{(D+1)/D}$.
Fig.~\ref{f2} shows the approach to this predicted scaling.
The exponents for linear chains (a) and fully-connected graphs (d) 
approach the predictions at very small sizes
compared to $2D$- and $3D$-lattices, which is due to the finite-size effects 
inherent in using Eq.~\ref{e4}.
 
The above analysis was for a very simple model, with only one mutation 
 present at any time. 
However, in realistic biological systems, multiple mutations can occur 
and interact. 
To investigate these effects, we have 
modified the model by allowing offspring not to be 
 perfect clones of their parents. 
In such a model,
the network topology has a 
significant impact on the evolved, equilibrium fitness of the population.
We also look at the effects of sexual reproduction, with Mendelian
recombination rules, and demonstrate that network topology also affects sexual
populations, but in a different way to asexual ones. 

\begin{figure}[ht]
\includegraphics[angle=270, width=0.49\textwidth]{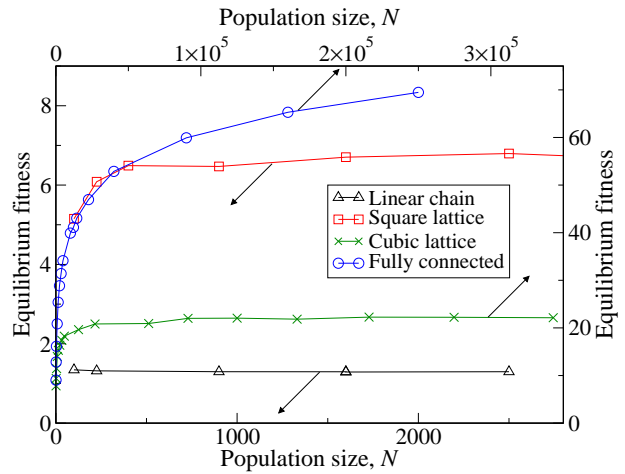}
\caption{
\label{f3}(Color online) 
Relationship of equilibrium fitness to population size, $N$, and network topology in an asexual population for $\mu=0.4$, $\Delta_0=1$ and $p=0.1$ 
(cf. Eq.~(\ref{e6})). 
Data were averaged over 50 realizations and 
errors are smaller than the symbols.
Lines are guides to the eye only.
The linear-chain and square-lattice systems use the left and bottom axes; the cubic-lattice and fully-connected systems use the top and right axes.
Initial condition: all fitnesses set to unity.
The numbers measuring equilibrium fitness are not direct measures of biological reproduction rates, being scaled by $\Delta_0$.
}
\end{figure}

In the asexual model with multiple mutations, 
each time that an offspring is produced, 
it is subject to a possible mutation.
Mutations occur with a probability, $\mu$, 
and increase the fitness with probability $p$,
or decrease it with probability $1-p$.
The fitness increment/decrement is the same, $\Delta_0$,
i.e. offspring are produced from clones of their parents, 
but with a stochastic change in their fitness 
$r_{\text{offspring}}=r_{\text{parent}}+\Delta$. 
The value of $\Delta$ is drawn from a probability distribution function,
$\rho(\Delta)$, of the form, 
\begin{equation}
\rho(\Delta)= 
(1-\mu)\delta(\Delta) + 
\mu[(1-p) \delta(\Delta + \Delta_0)+ p\delta(\Delta - \Delta_0)]~, 
\label{e6}
\end{equation}
(with $\Delta_0>0$). 
In this model (as we have observed numerically), 
the population reaches an equilibrium fitness, caused by mutation-selection
balance \cite{Burger:book}, proportional to $\Delta_0$  
provided that the initial dynamics do not lead to extinction 
(in which all individuals' fitnesses reach zero) and that $p<0.5$ 
(with more positive mutations than negative mutations, the fitness tends to infinity). 
In the fully-connected system, this is the model studied in Ref.~\cite{Ridgway_98}.
Fig.~\ref{f3} shows that 
the equilibrium fitness of asexual populations significantly depends on 
the dimensionality of the network as well as  on   
the population size \cite{Lande_95},  with higher dimensional systems exhibiting 
greater equilibrium fitnesses. 
The networks found (Fig.~\ref{f1}) to have a slower
spread of mutants 
lead to a lower fitness in the model with multiple mutations.
The slower spread means that individuals compete with closely related
neighbours and when 
a fit individual reproduces, it is more likely to replace another fit individual.
It can be seen from Fig.~\ref{f3} that the fitness increase with $N$
saturates for the spatial systems.  
For fully-connected systems of up to 250,000 nodes, no saturation was observed
(cf. Ref.~\cite{Ridgway_98}).

\begin{figure}[ht]
\includegraphics[angle=270, width=0.49\textwidth]{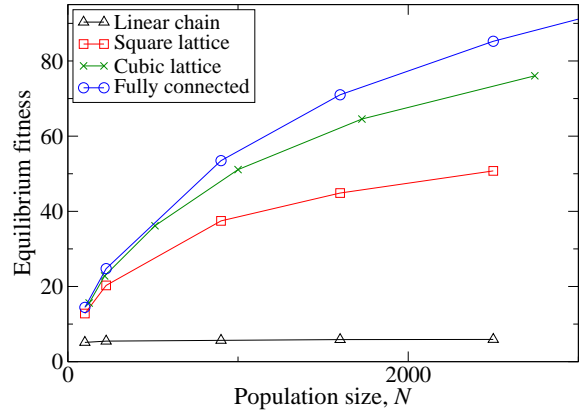}
\caption{
\label{f4}(Color online) 
Relationship of equilibrium fitness to population size, $N$, and network topology in a sexual population for $\mu=0.04$, $\Delta_0=1$ and $p=0.1$ 
(c.f. Eq.~(\ref{e6})) and $G=10$ (so that $\mu G =0.4$ has the same value as $\mu$ in Fig.~\ref{f3}). 
The data were averaged over 50 realizations and 
the errors are smaller than the symbols used.
Lines are guides to the eye only. 
Initial condition: all gene quality factors set to unity.
}
\end{figure}

To study a sexual population, 
it is necessary to introduce multiple genes per individual, allowing recombination.
We study an additive model 
(see e.g. 
\cite{Maynard_Smith_68}) in which each individual has $G$ genes.
Each gene is characterised by a number (quality factor),
and the fitness of the individual is the sum of these quality factors.
When an individual is chosen for reproduction, a partner is selected 
at random from the nodes connected to it 
and the offspring is produced by a Mendelian shuffling of the genes.
The above criteria define hermaphrodite haploids 
(bisexual individuals with a single copy of each gene).
Mutations are applied as in the asexual system, but now each gene
is subject to a possible mutation.
In Fig.~\ref{f4}, the equilibrium fitness of the sexual population 
\textit{versus} 
$N$ is shown for different network topologies, 
demonstrating that the dimensionality
of the network also affects the fitness of sexual populations, with higher
dimensionality networks again leading to increased fitnesses.
However, the effect is not as large as for the asexual
populations. 
A saturation equilibrium fitness with $N$ was not observed in the $2D$-, $3D$-lattices or 
fully-connected systems for $N \le 250,000$ nodes.
The sexual populations were found to have higher equilibrium fitnesses than 
the asexual fitnesses for the parameter range $G \gtrsim 10, \mu \gtrsim 0.01, p \lesssim 0.1, N \gtrsim 200$.
The general principle that the advantage of sex is greater (and the fitness is
lower) in lattices than in fully-connected graphs holds over a range of variables.

It should be noted that in an asexual system created by suppressing the recombination in a sexual system (i.e. an asexual population with genes) the fitness depends only on the product $\mu G$.
The populations in Figs~\ref{f3}-\ref{f4} are therefore directly comparable, and the differences are due to the presence or absence of recombination.
The 'slower' networks still see the fitness disadvantage, as with the asexual system.
However, the correlations in the population fitness 
(a fit neighbour being surrounded by other fit individuals) 
now give a benefit as well as a disadvantage - a fit individual is more 
likely to mate with a fit individual as well as to replace one with her offspring.
The sexual process may also reduce the degree of spatial frustration, 
as genes can pass to second-nearest neighbours, potentially speeding up the spread of mutations.
A larger advantage for sex does not necessarily mean, however, 
that the establishment/maintenance of sex would occur more 
readily \cite{Paley_07:1} in a spatial network than in the fully connected systems 
usually studied (to be discussed elsewhere).

To conclude, we have demonstrated that the spatial nature of populations 
(dimensionality of the population network) is of significant importance 
for the dynamics of evolution. 
The following effects have been found for populations 
defined on networks of higher dimensionality (e.g. fully-connected graph as
compared to a square lattice): 
(i) faster spread of a mutation, 
(ii) greater values of equilibrium fitness in both sexual and asexual 
populations, in a model in which mutations do not have time to fixate
before the next mutation is introduced into the system.
Whilst both sexual and asexual populations have increased fitnesses for
higher dimensional networks, the effect of topology on each is markedly different. 

CJP thanks the EPSRC for financial support. 



\bibliography{archive_cjp2}

\section{Supplementary information}

\section{The discrete origins of the master equation}

Each individual on the graph has a fitness, 1 (wild-type) or $r$ (mutant) in this part of the paper, associated with it.
On each turn, an individual is selected for reproduction, with a probability proportional to its fitness.
A clone of the selected individual is placed onto one of the nodes connected to it by the network.
The individual that previously occupied the node is replaced, conserving the population size, $N$.

For the spread of mutants, the important events are those that occur at the interface between mutants and non-mutants.
At a given connection between a mutant and a non-mutant node, the probability that an event occurs at one of the nodes in a turn is:
\begin{equation}
p_{conn} = \frac{1+r}{mr+(N-m)} ~.
\end{equation}
At a given interface link there must be a wild type (fitness of unity) and mutant individual (r).  The probability of selecting one of these is normalised by the total mutant ($m \times r$) and wild type ($[N-m]\times 1$) fitness.
Given that an event has happened at a node linked to a particular connection between mutant and non-mutant, the probability that the offspring is placed across this connection is $1/Z$ where $Z$ is the connectivity of each node.
In total, there are $A$ links between mutants and non-mutants.
The probability that, in a given turn, an event happens across the boundary is: 
\begin{equation}
p_{boundary} = \frac{A}{Z}\frac{1+r}{mr+(N-m)} ~.
\end{equation}

Therefore, the probability that, in a given turn, the population of mutants increases by one is:
\begin{equation}
p_{\uparrow} = \frac{A}{Z}\frac{1+r}{mr+(N-m)} \frac{r}{1+r} ~.
\end{equation}
The probability that the mutant population shrinks by one is
\begin{equation}
p_{\downarrow} = \frac{A}{Z}\frac{1+r}{mr+(N-m)} \frac{1}{1+r} ~.
\end{equation}

This can be written as a discrete master equation, where the probability that there are $m$ mutants in the system at time $t$ is given by:
\begin{equation}
P_{m,t} = p_{\uparrow}P_{m-1,t-1} + p_{\downarrow}P_{m+1,t-1} + (1-p_{\uparrow}-p_{\downarrow})P_{m,t-1} ~.
\end{equation}

Alternatively, the continuum, deterministic approximation can be taken, where the probability of increase minus the probability of decrease per time-step in mutant population leads to an average rate of increase (valid for $1<<m<<N$):
\begin{equation}
\frac{dm}{dt}=R_{\uparrow}-R_{\downarrow} = \frac{A}{Z}\frac{r-1}{mr+(N-m)} ~,
\label{eq:master}
\end{equation}
where $R_{\uparrow(\downarrow)}= p_{\uparrow(\downarrow)}/\Delta t$ 
(with $\Delta t=1$ being the simulation time step length).
This is equation (1) in \cite{Paley_06:arXiv}.

\section{Derivation of the relation between the number of mutants and time on a lattice}

We use the approximation,
\begin{equation}
A = \beta m^{\gamma}~, 
\label{e4}
\end{equation}
and place this into equation (\ref{eq:master}) before integrating.
\begin{equation}
\int_{1}^{m}\frac{mr+(N-m)}{m^\gamma}\mathrm{d}m=\int_{0}^{t}\frac{\beta}{Z}(r-1)\mathrm{d}t~.
\end{equation}
This integral can be solved to give:
\begin{equation}
(r-1)\frac{m^{2-\gamma}}{2-\gamma} + N\frac{m^{1-\gamma}}{1-\gamma} -(r-1)\frac{1}{2-\gamma} - N\frac{1}{1-\gamma} =\frac{\beta}{Z}(r-1)t~,
\end{equation}
which can be re-arranged to give equation (5) in \cite{Paley_06:arXiv}.

\section{Explanation of the equation describing the introduction of new mutations}

In the second part of the paper, a model is introduced in which parents do not necessarily produce perfect offspring.
The offspring's fitness is related to that of the parent by the following formula:
\begin{equation}
r_{\text{offspring}}=r_{\text{parent}}+\Delta ~.
\end{equation}
The difference in the fitnesses, $\Delta$, is taken from a probability distribution,
\begin{equation}
\rho(\Delta)= 
(1-\mu)\delta(\Delta) + 
\mu (1-p) \delta(\Delta + \Delta_0)+\mu p\delta(\Delta - \Delta_0)~. 
\label{e6}
\end{equation}
This distribution consists of three weighted delta-functions, representing the three discrete possibilities.
With probability $\mu (1-p)$ (the probability that a mutation occurs multiplied by the probability that it is deleterious), the offspring will have a fitness of $\Delta_0$ less than that of their parent ($\Delta=-\Delta_0$).
With probability $\mu p$, the offspring will have a fitness of $\Delta_0$ more than that of their parent, representing an advantageous mutation.
With probability $1-\mu$, no mutation will have occurred and the offspring will have the same fitness as the parent.

\end{document}